\documentclass[12pt,preprint]{aastex}
\long\def\beginpgfgraphicnamed#1#2\endpgfgraphicnamed{\resizebox{\hsize}{!}{\includegraphics{#1}}}

\shorttitle{A Precessing Jet in the CH Cyg}

\shortauthors{Karovska, Gaetz, Carilli, Hack, Raymond \& Lee}
\begin{document}

\title{A Precessing Jet in the CH Cyg Symbiotic System}

\author{Margarita Karovska\altaffilmark{1}}
\author{Terrance J. Gaetz\altaffilmark{1}}
\author{Christopher L. Carilli\altaffilmark{2}}
\author{Warren Hack\altaffilmark{3}}
\author{John C. Raymond\altaffilmark{1}}
\author{Nicholas P. Lee\altaffilmark{1}}

\altaffiltext{1}{Smithsonian Astrophysical Observatory, 60 Garden
  Street, Cambridge, MA 02138 \\
  Corresponding author: mkarovska@cfa.harvard.edu}
\altaffiltext{2}{National Radio Astronomy Observatory, PO Box O, Socorro, NM 87801}
\altaffiltext{3}{Space Telescope Science Institute}

\begin{abstract}

{Jets have been
detected in only a few symbiotic binaries to date, and CH Cyg is one
of them.
In 2001, a non-relativistic jet was detected in CH Cyg for the first time in X-rays.
We carried out coordinated {\it Chandra}, {\it HST}, and {\it VLA}
observations in
2008 to study the propagation of this
jet and its interaction with the circumbinary 
medium.  We detected the jet with {\it Chandra} and {\it HST} and
determined that the apex
has expanded to the South
from $\sim$300 AU to $\sim$1400 AU, with the shock front propagating
with velocity $<$100 km s$^{-1}$. The shock front has
significantly slowed down since 2001.
Unexpectedly, we also discovered a powerful jet in the NE-SW 
direction,
in the X-ray, optical and radio.
This jet has a multi-component structure, including an inner jet and a
counter-jet at $\sim$170 AU,
and a SW component ending in several clumps extending out 
to $\sim$750 AU.
The structure of the jet and the curvature of the outer portion of the SW
jet suggest an episodically powered precessing jet, or a
continuous precessing
jet
with occasional mass ejections or pulses.
We carried out detailed spatial mapping of the
X-ray emission and correlation with the optical and radio emission.
X-ray spectra were extracted of the central source, inner
NE counter jet, and 
the brightest clump at a distance of $\sim$500 AU from the central source.
We discuss the initial results of our analyses, including 
the multi-component spectral fitting 
of the jet-components and of the
central source.}

\end{abstract}


\section{INTRODUCTION}

Symbiotic systems are accreting binaries
which are potential 
progenitors of a fraction of Pre-Planetary and Planetary Nebulae, 
and of a fraction of
SN type Ia (e.g. Chugai $\&$ Yungelson 2004).
CH Cyg is a symbiotic containing mass-losing M6-7 III giant and
an accreting white dwarf (e.g. Hinkle et al. 2009).
At only $\approx$250 pc (Perryman et al. 1997),
CH Cyg is one of very few interacting binaries
close enough 
for detailed multiwavelength spatial and spectral studies of
the close circumbinary environment (e.g. Corradi {\it et al.} 2001, Eyres {\it et al.} 2002,
Crocker {\it et al.} 2002, Ezuka {\it et al.} 1998, Mukai {\it et al.}
2007, Scopal{\ et al 2007}, Pedretti {\it et al.}, 2009).
In 1984 a powerful radio jet was detected in CH Cyg (Taylor at al. 1986)
following a sudden dimming of the V magnitude.
Similar radio jet-activity was detected in 1998 following 
another sudden V magnitude dimming in 1996 (Karovska {\it et al.}
1998).

CH Cyg, one of a very few symbiotic systems with
jet activity also seen in X-rays, was directly imaged for the first time 
in 2001 using {\it Chandra} (Galloway and Sokoloski, 2004,
Karovska {\it et al.} 2007).
Karovska {\it et al.} (2007) detected multiple components 
in {\it Chandra} images obtained in March 2001, including   
a loop-like structure associated with an expanding jet in the soft
($<$2 keV) image extending $\approx$1.5'' (375 AU) to the south of the central
source.  It was also visible in the
{\it VLA} and {\it HST} images.  The emission 
was consistent with optically thin thermal X-ray
emission from shocks between the jet and
the dense circumbinary material.

In June 2008 we carried out follow-up {\it Chandra}, {\it HST}, and {\it VLA} observations of
CH Cyg to monitor the propagation of the 2001 X-ray jet.
We detected the 2001 jet with {\it Chandra} and {\it HST} as 
it has expanded in the circumbinary
environment, and discovered powerful, $\it new$ jet activity
at X-ray, radio, and optical
wavelengths. In \S2 we describe the
observations and analyses, and the initial results are summarized in \S3.
We discuss the results in  \S4.

\section{OBSERVATIONS AND ANALYSIS}


{\it Chandra} observations of CH Cyg were carried out 
on 2008 June 8-10 (ObsIDs 8972, 9867, and 9868) obtaining
$\sim$80 ks of data (Table 1) with ACIS-S3 
(Weisskopf {\it et al.} 2002). 
We analyzed the observations using {\it CIAO}\footnote{Chandra Interactive Analysis of Observations
system package (http://cxc.harvard.edu/ciao)} 4.1 and CalDB version 4.1.1
We merged all ObsIDs, using
ObsID 9867 as the coordinate reference.
Source and background spectra were extracted 
by
standard methods following {\it CIAO} threads.
The spatial
distribution of the X-ray emission was explored using sub-pixelated images 
(0.1''/pixel) and PSF simulations and deconvolution
techniques, taking  advantage of the telescope dither,
which provides access to smaller 
spatial scales than the ACIS pixel (0.492") as
the image moves across the detector pixels. 
Similar techniques were applied for resolving the CH Cyg jet in 2001
(Karovska {\it et al.} 2007),
the Mira AB symbiotic binary
system (Karovska {\it et al.} 2005), substructures
in NGC6240 (Esch et al 2004),
and SN1987A (e.g. Burrows {\it et al.} 2000, Park {\it et
al.} 
2002).
We applied a statistical deconvolution technique,
 EMC2, 
specifically applicable to low-counts Poisson data (Esch {\it et al.}
2004; Karovska {\it et al.} 2005, 2007).
The deconvolution was performed using simulated Chandra
ChaRT\footnote{http://asc.harvard.edu/chart/index.html} PSFs
(Karovska {\it et al.} 2003).

We carried out {\it HST} observations on 2008 June 9 using WFPC2 with the
filters listed in Table 1.
A 4-point dither pattern was used in each filter to
better sample the PSF and for improved
cosmic-ray rejection. 
The central region of the CH Cyg system was located at the edge of the
WF2 chip which has a resolution of 0.0996" per pixel. 
All images were calibrated using the standard pipeline calibrations
({\it vis} Baggett et al., 2002).
The four exposures for each filter were combined
together at a plate-scale of 0.05" per pixel using MultiDrizzle 
(Koekemoer et al. 2002). 
Model PSFs derived from $\it Tiny Tim$ ({http://www.stsci.edu/software/tinytim/}) allowed identification of 
features related to the PSF itself.

On 2008 October 4, we obtained {\it VLA} observations (A
configuration) at 5 GHz, 8 GHz (3 hours
each, 2 hours on source), and at 1.4 GHz (1 hour).  
At 5 GHz we detected a total flux density from
the source of 4.6 mJy, with a peak surface brightness of 1.33 mJy
beam$^{-1}$.  The peak surface brightness at 8 GHz and 1.4 GHz was
1.71 mJy beam$^{-1}$, and  1.4 mJy beam$^{-1}$, respectively.
We achieved a resolution of
FWHM = $0.55\arcsec \times 0.42\arcsec$ 
at 5 GHz,
$0.30\arcsec \times 0.25$ 
at 8 GHz, and $1.2\arcsec \times 1.0\arcsec$ 
at 1.4 GHz, reaching an RMS
sensitivity of 19 $\mu$Jy at 5 GHz and
21 $\mu$Jy at 8 GHz, and 60$ \mu$Jy at 1.4 GHz.
The data were analyzed using standard
calibration and analysis procedures, including photometry and spectral
diagnostics of detected components (as in Karovska et al. 2007).

\section{INITIAL RESULTS}

The 2008 observations showed
a central source and two jets; the ``old'' (2001) jet in the SE direction,
and a new jet in the SW/NE direction.
In the following we present the initial results from the {\it
Chandra}, {\it HST} and {\it VLA} imaging and X-ray spectroscopy.

{\it 1. Imaging}

{\it Old Jet}

Figure 1 (left panel) displays the 2008 {\it Chandra} image
showing the ``old'' SE jet; a loop-like structure extending south of the central
source at PA =175$^\circ$ to a distance of 5.6'' ($\sim$1400 AU).
About 90 counts were detected,
with a hardness ratio ( hereafter HR, counts 0.2-2 keV/
counts 2-8 keV).

The ``old'' SE jet was also detected in the 2008 {\it HST} H$\alpha$ image, as shown in
Figure 1 (right panel), at about the same position as in the {\it Chandra} image.  
The color levels emphasize the extended loop structure rather then inner
structures associated with the new jet.
below.
The loop morphology is very similar in the {\it HST} and {\it Chandra} images and
structures appear co-spatial. However,
there may be a spatial displacement between the
structures, limited by the size of the ACIS-S pixel size of 0.492''. 
To show the expansion of the jet since 2001 we overlaid on the 2008
{\it Chandra} and {\it HST} images the contours of the X-ray image obtained in
2001 (ObsID 1904) showing the ``old'' jet loop-like structure at
1.5''.

The loop structure is likely due to a shock front propagating through
the extended circumbinary material.
The loop is clearly detected only in H$\alpha$ and [S II],
indicating emission from a shock slower than 100 $\rm km~s^{-1}$.
Assuming a linear expansion since 2001, the average velocity of the jet
would be $\approx$  700 $\rm km~s^{-1}$.
However, if the 100 $\rm km~s^{-1}$ shock is representative of the current expansion
velocity of the jet apex, then the initial
speed of the jet may have been much higher, exceeding 1000 $\rm
km~s^{-1}$, and the jet has
slowed down during its propagation in
the circumbinary medium.

{\it New Jet}

Figure 2 displays the F502N (5007 \r{A} [OIII]) 
{\it HST} image showing the central
source region (C0) and the components of the new jet.
An inner SW jet and a NE counter-jet (C2 and C1,
respectively),
both extend to about 1'' from the central source, and
a fainter narrow structure extends to the SW. It reaches a
bright knot (C3) at $\sim$2'' with P.A. of $\sim$215$^{\circ}$ ,  then
expands into a hook-like structure.  The hook-like structure contains
several bright clumps as it curves $\sim$2.5-3'' toward
the West.
The bright C3 clump dominates the emission
from the hook-like structure.  It is observed clearly at F502N ([OIII])
and
in the other {\it HST} filters except for F547M and F673N ([S II]).
The absence of emission from the lowest ionization material suggests
photoionized gas or a shock in which recombination
is incomplete (Raymond {\it et al.} 1988).

Several of the structures in the {\it HST} images were also 
detected in the {\it Chandra} images.
These include the central source, the SW inner jet, the
NE inner counter-jet, and the clump C3. 
A bridge-like structure is seen between the inner jet and
the clump C3, and possibly emission associated with C4 and C5.
The structures in the new jet are bright enough for sub-pixel
analysis using the information on the photon
positions
({\it vis} Karovska {\it et al.} 2005). 
We created energy-filtered images in the soft 0.2-2keV
band, medium 2-4 keV band, hard 4-6 keV band and in the 
6-7 keV band (encompassing the strong
6.67 keV Fe line).
The NE inner counter-jet (C1) is clearly detected in the {\it Chandra}
images, with a hardness ratio
HR=0.15.
The brightness of C3, which dominates the soft X-ray
emission from the system below 2 keV, drops quickly at
energies $>$2 keV. The HR is
0.05. 
Figure 3 shows the deconvolved {\it Chandra} 0.2-2 keV image,
overlaid with contours of the  6-7 keV image. 
For comparison we also show the overlaid
contours of the F502N {\it HST} image (Fig. 3 left panel) and of the {\it VLA}
5 GHz image (Fig 3 right panel).
At the $\approx$0.2'' resolution, the {\it Chandra} emission 
appears co-spatial with the {\it HST}
F502N emission in C0, C1, and C3.

The central source, the SW inner jet, the NE inner counter-jet, and a 
portion of the hook-like  structure are also
seen 
in the {\it VLA} 1.4 GHz, 5 GHz, and 8 GHz maps. There is a faint elongated blob
in the 5 GHz map corresponding to the C3 region (Figure 3 right
panel). 
However, 
no significant radio emission is seen at C3 at 8 GHz. 
Very faint emission close to the C4, and C5 knots is detected in
the 5 GHz map.

The radio jet is clearly resolved in both width and length 
($\sim$3" long, 1" wide).
The integrated emission shows roughly flat spectrum from 1.4 to 8 GHz 
consistent with optically thin free-free emission.
The southern loop (leaving out the core + core-jet on a scale of 
$<$ 0.5") has an integrated radio flux of 1.4 mJy at 5 GHz and 2.0 mJy at
8 GHz. 
From the radio flux density and morphology, we  estimate the 
following mean values over the source:
ionized gas density $\sim$ 1.5$\times$ 10$^{4}$ cm$^{^-3}$ (assumed temperature
$\sim$ 1$\times$ 10$^{4}$ K),
pressure $\sim$ 3$\times$ 10$^{8}$ K cm$^{^-3}$  (or 4$\times$ 10$^{-8}$ dyn cm$^{^-2}$).
We find the HII mass to be 5$\times$ 10$^{-6}$ M$\sun$, and the emission
measure to be 6.7$\times$ 10$^{5}$ pc cm$^{-6}$.

{\it The Central Source}

A bright central source is detected in all images.  
It appears to be resolved, and possibly elongated in
the direction perpendicular to the new jet.
The central source is fainter than C3 in the soft 0.2-2 keV band, 
but appears bright at harder wavelengths,
including the Fe emission at 6-7 keV.
We also detected variable X-ray emission
from the central source.
Further analysis is underway
in determining the nature of the extended emission and the
variability.

{\it 2. X-ray Spectroscopy}

Figure 4 shows the spectra extracted in three
main regions, non-overlapping ellipses centered on: 
 (1) the central
C0 source, (2) the inner NE counter-jet
source C1, and (3) the bright SW clump C3 (see Fig.3).
The spectra were grouped 
to a minimum of 10
counts per bin, and the fits were performed using
{\tt Xspec} 12.5.0\footnote{Xspec is an X-ray spectral fitting package 
(http://heasarc.gsfc.nasa.gov/docs/xanadu/xspec/)} using the $\chi{^2}$
statistic and 
``churazov'' weighting.
The spectra for the three extraction regions differ markedly.
All three regions show significant soft emission ($\la 2$\,keV).

In C3, the spectrum falls off sharply above $\sim 2\,\mathrm{keV}$,
characteristic of a soft thermal spectrum, but also has significant,
very soft emission ($\sim 0.2$--0.4\,keV).  We fit this spectrum
with a combination of absorbed ({\tt tbabs}) thermal ({\tt vapec})
models.
C0 and C1, also show significant hard emission
and prominent Fe-K lines.  The hard emission is very strong in region C0,
and relatively weaker in region C1.  Because the source regions are
very close in projection, some emission from each region
(from the wings of the PSFs) 
contaminates the
extracted spectrum in the other.
We assume that the hard spectrum originates entirely
from region C0 and the soft emission entirely from region C1,
but that smearing by the PSF wings
mixes fluxes from the two sources.
We fit the spectra simultaneously.  Each component is modeled as
a combination of absorbed thermal components, with a scaled fraction
of each spectrum contributing to the other spectrum.  
In the fit to
region C0, the soft components are scaled by $\sim 0.5$, while
in region C1, the hard component (and the Gaussians) are scaled by
$\sim 0.1$.

The soft components were fitted by relatively unabsorbed soft thermal models
($N_H\sim$ 0, $kT_\mathit{e} \approx 0.2$ keV, and
$N_H\approx 2\times 10^{21}\,\mathrm{cm}^{-2}$, 
$kT_\mathit{e} \approx 0.64\, \mathrm{keV}$).  
To better account for the intermediate energy flux, and for the hard component, we used a
partial covering absorber (${\tt pcfabs}$) model.  The resulting
model for the hard component required a hot 
($kT_\mathit{e} = 7.2^{+1.4}_{-1} \mathrm{keV}$) very heavily absorbed 
($N_H \sim 4.1\times 10^{23}\,\mathrm{cm}^{-2}$) thermal model with
$\approx 99$\% coverage fraction.  The leakage with the partial
covering model significantly improved the fit at intermediate energies,
but a systematic $\sim 2-4\,\mathrm{keV}$ excess remains for
C0; this will be explored in a follow up paper.
The large covering factor suggests that K$\alpha$ emission from low
ionization states should be present, and indeed features at 6.4keV and
$1.77\pm0.03$ keV corresponding to Fe K$\alpha$ and Si K$\alpha$ are detected.
Although the model for the hard component produced significant Fe-K emission,
the very large residuals indicated that at least one additional line 
component is needed.  This was implemented as a ``zero-width Gaussian''
(i.e., a Gaussian with negligible width compared to the instrumental
resolution) at
6.4 keV, consistent with neutral or low ionization Fe.  In the
C0 spectrum, a prominent line-like feature at $1.77\pm0.03$ keV
is consistent with Si-K emission from neutral or low-ionization Si.

The C3 X-ray spectrum appears to be mainly from a thermal plasma, with
O, Ne, and Fe-L line emission (0.5--1 keV), and a prominent
Mg XI line feature at 1.34 keV.  We fit the spectrum with absorbed 
({\tt tbabs}) thermal ({\tt vapec}) plasma models.  In order 
to match the spectrum at very soft ($\lesssim 0.4\,\mathrm{keV}$),
intermediate (1-2\,keV) and harder ($\gtrsim 2\,\mathrm{keV}$) energies,
we needed three thermal components with $kT \approx 0.17\,\mathrm{keV}$,
$kT \approx 0.27\,keV$, and $kT \approx 1.77\,\mathrm{keV}$;  
the fitted absorption was effectively zero.  Fits with 
pure solar abundances did not fit well (particularly the Mg XI 
line feature at 1.34 keV), so the abundances of O, Ne, Mg, and 
Fe (= Ni) were allowed to vary (but with the same values for 
each of the thermal components).  The fitted [Fe] and [Ne] abundances
were effectively solar (Anders \& Grevesse 1989 abundances); 
[O] was low ($\sim 0.4$), while [Mg] was very high ($\sim 3.5$).
The 
resulting residuals suggested a line-like feature at about 
1.66 keV; it may be due to excess Mg XII emission, 
beyond what is predicted by the three component thermal model.

\section{DISCUSSION}

The hard emission component from the central region, C0,
observed in 2008 shows a striking resemblance in both
temperature and high absorbing column to the hard emission
observed in 2001 (Galloway \& Sokoloski 2004).
This indicates that the hard emission originates close
to the accretor and may be associated with the 
boundary layer of the
accretion disk or a bright spot (Kennea {\it at al.} 2009).
The emission from this region is heavily absorbed by a partial covering
absorber, with a small ``leakage'' accounting for some soft emission.
The absorbing material could be related to the ejecta close to the
accreting source or to the accretion disk, and may be associated with the
material in the inner jet.
The soft emission near the center at C1 is generally similar
to the emission at C3, but it is more difficult to characterize
because of the blending with radiation from C0.  We suggest
that it arises from the same mechanism, shock waves driven
into the ambient gas by the jet.

The new jet activity
is likely a result of an ejection following the strong
dimming of the light curve of CH Cyg in the second half of 2006 (eg. Skopal et al. 2007).
Taranova and Shenavrin (2007) found evidence for dust
formation in CH Cyg which
resulted from 
a mass ejection in
the close circumbinary region associated with the 2006 light dimming
event.
The light curve is very similar to the 1984 and 1996 dimming events 
after which
jets were detected in radio and optical wavelengths.
Assuming that the ``new'' jet was ejected in the second half of 2006, and a
linear expansion, the velocity 
could be over 1500 km/s.

The SW inner jet and the NE inner counter-jet are located within 1'' ($\sim$250 AU) from the central
source, which is significantly smaller than the 
$\sim$3$^\prime$$^\prime$ (750 AU) SW portion of the jet. It is possible that this is the
inner part of a continuous precessing jet, since
a similar
elongated structure, extending to  $\sim$ 0.7'' at PA =330$^\circ$,
and fainter structure to the south at about same distance from
the central source at PA = 150$^\circ$ were 
detected in the {\it Chandra} images from 2001 ({\it vis} Fig. 5 in Karovska
{\it et al.} 2007). 

The inner jet and the inner counter-jet may be running into a shell of dust at $\sim$1'' from
the central source, which was detected 
by Biller {\it et al.} (2006). Biller {\it et al}. showed that the
dust shell is significantly denser to
the N, which could account for the difference in brightness of the C1
and C2 structures, if they are indeed a result of an interaction of
the jet with the
dust shell. 
After encountering the dust shell, the inner jet (C2) is then
continuing to the SW,	 
while the inner counter-jet (C1) is 
slowed down significantly by the denser shell to the N.

The curved appearance of the outer portion of the SW jet 
and the observed clumps is consistent with
an episodically powered precessing jet or a continuous jet
with occasional mass ejections, or pulses
($\it vis$ Crocker et al. 2002; Stute and Sahai 2007).
The clumps, including the dominant C3 structure, may be shocked ejecta
interacting with the circumbinary material (eg. with mass loss from the red
giant and/or with previously ejected material).  We note that
similar precessing jet-like outflows have been found in Pre-Planetary
and Planetary nebulae (e.g. Sahai and Trauger; 1998; Soker, 2002;
Sahai {\it  et al}. 2005), a fraction of which possibly evolve 
from
symbiotic binaries like CH Cyg.

The fits to the C3 X-ray spectrum give temperatures corresponding to 400 to 1200 $\rm km~s^{-1}$ 
shocks.  The spectral extraction region
corresponds to a length scale of $4.4\times 10^{15}$ cm.  If the
line of sight depth of the region is comparable in size, the
X-ray fit normalizations would imply densities of 
1200, 1000, and 450
$\mathrm{cm}^{-3}$ for the soft, medium, and hard
components, respectively.  The total internal energy would then be
$\sim 5\times 10^{41}\,\mathrm{erg}$.

In summary, we presented the initial results from the analysis of
the 2008 {\it Chandra}, {\it HST}, and {\it VLA} observations of the 
CH Cyg system
including the detection of the ''old''  2001 jet which has expanded to
a distance of 1400 AU to the SE, and a discovery of a new jet extending to
$\sim$750AU to the SW. We discussed the
initial results from the
multi-wavelength imaging
and X-ray spectroscopy.

\acknowledgments

We 
are grateful to AAVSO for the CH Cyg light curve.
We thank Dr. Miller Goss for useful discussions.
MK, TJG, and NL are members of the {\it Chandra} X-ray Center,
which is operated by the Smithsonian Astrophysical Observatory under
NASA Contract NAS8-03060. 
The VLA of the NRAO, is a facility of the NSF
operated by the Associated Universities, Inc.
Based on observations (program No. 11342) made with the NASA/ESA HST, 
obtained at the STScI, which is operated by 
the AURA, Inc.

\begin{table}[ht]
  \centering
  \begin{tabular}{l}
    \hline\hline

\emph{Chandra} ACIS-S: \\
\begin{tabular}{c|c}
    ObsID & Exposure Time (ks) \\ \hline
    8972  & 26.0 \\
    9867  & 27.5 \\
    9868  & 26.5 \\
\end{tabular}

\\
\\

\emph{HST} WFPC2: \\
\begin{tabular}{cl|c}
    Filter &                       & Exposure Time (s) \\ \hline
    F336W & Str\"omgren U & 400              \\
    F375N & [OII] & 800                      \\
    F437N & [OIII] & 2000                   \\
    F502N & [OIII] & 400                    \\
    F547M & Str\"omgren Y           & 64                \\
    F656N & H$\alpha$              & 400               \\
    F673N & [SII]       & 800               \\
\end{tabular}

\\
\\

\emph{VLA} A-Configuration: \\
\begin{tabular}{c|c}
    Frequency (GHz) & Observing Time (hours) \\ \hline
    1.425           &       1                \\
    4.9             &       3                \\
    8.4             &       3                \\
\end{tabular}

\\

    \hline\hline
  \end{tabular}

  \caption{{\it Chandra}, {\it HST} and {\it VLA} Observations}
  \label{tab:obssummary}
\end{table}

\clearpage

\begin{figure}
\epsscale{.85}
\plotone{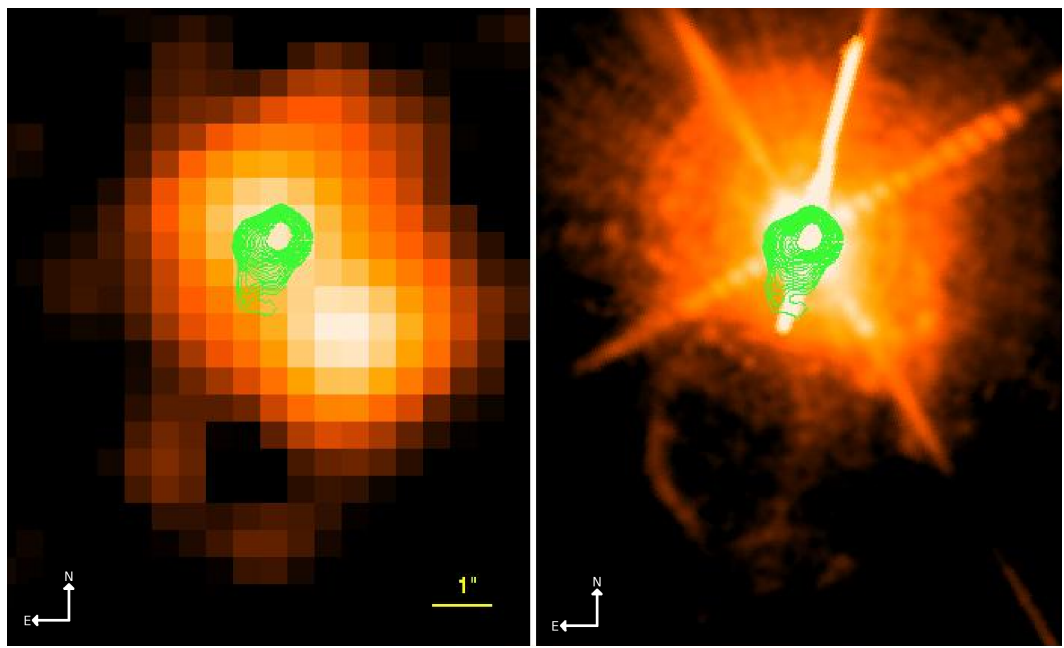}
\caption{
(left panel) 2008 {\it Chandra} image of the ``old'' jet at a distance
of 5.6'' SE from the central source. This jet was first detected in
2001 with {\it Chandra} at 1.5'' from the central source
(overlaid green contours). 
(right panel) 2008 {\it HST} F656N (H$\alpha$) image showing the
jet at a similar distance as in the 2008 {\it Chandra} image.
To show the expansion of the jet we overlaid contours (green) of the 2001
{\it Chandra} image. (The white linear features are artifacts due
to the ``bleeding'' of the chip.)}
\end{figure}

\begin{figure}
\epsscale{.85}
\plotone{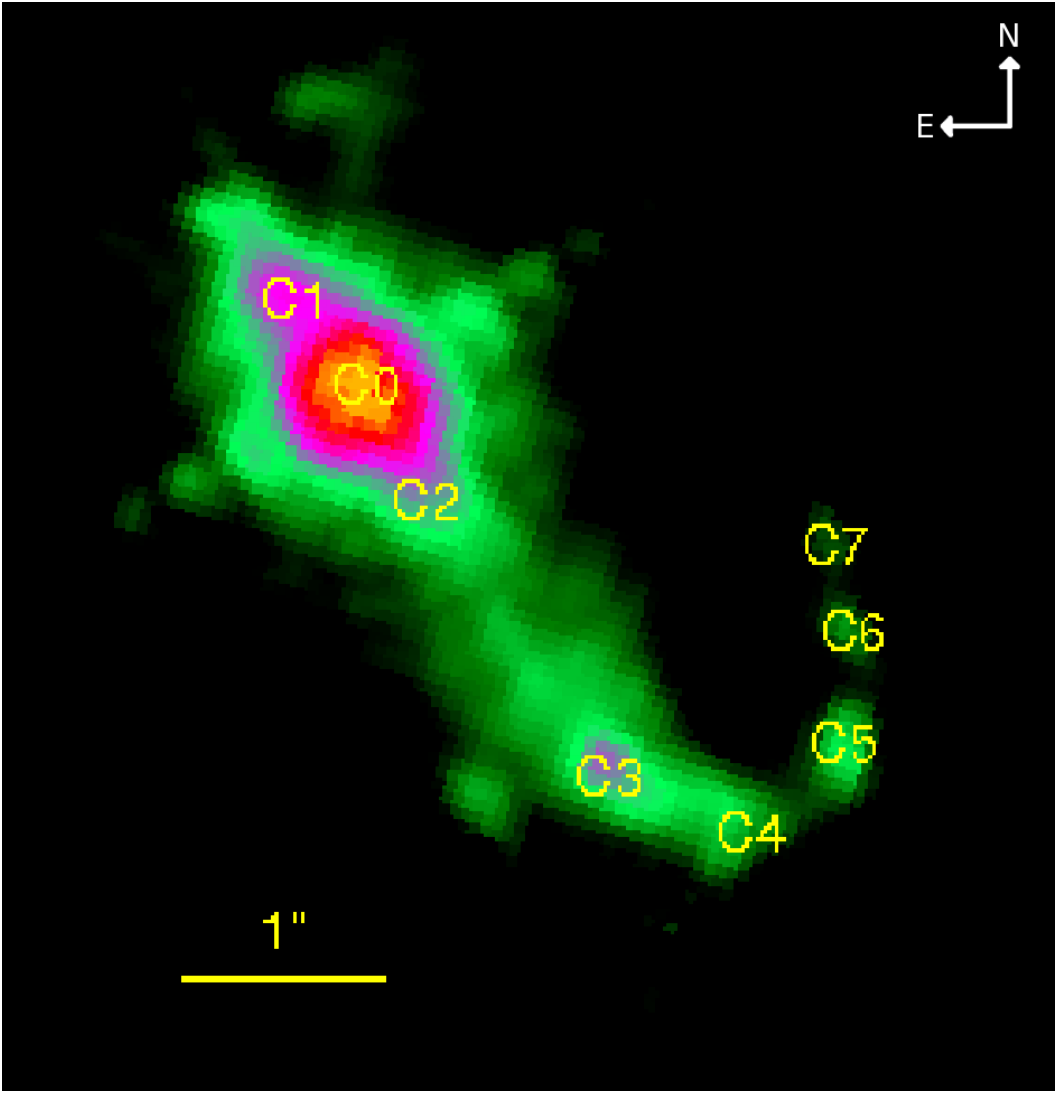}
\caption{
{\it HST} F502N image shows the
bright central source, the inner SW jet and NE counter-jet, both at about 1'' from
the central source, and a hook-like structure
to the S-W including several clumps (C3-C7) at a distance from 2-3''
from the central source.
A bright clump C3 is clearly visible at 2'' from the central source,
at a P.A. of $\sim$215$^{\circ}$.}
\end{figure}

\begin{figure}
\epsscale{.85}
\plotone{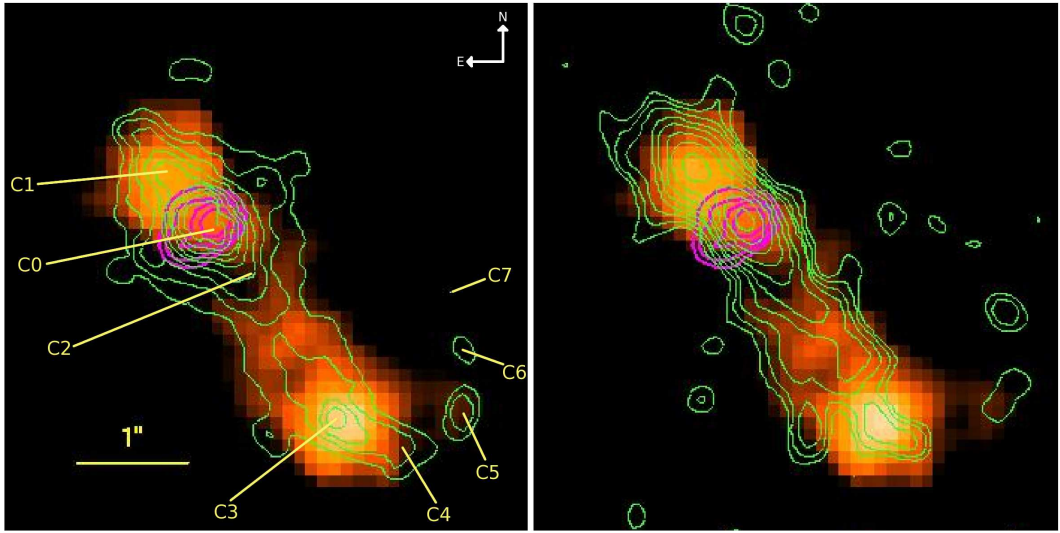}
\caption{
(left panel) Deconvolved soft (0.2-2 keV) {\it Chandra} image with overlaid
contours (magenta) of
the 6-7 keV image, and {\it HST}
502N contours (green).
(right panel). Deconvolved {\it Chandra} soft image with overlaid
contours of
the 6-7 keV X-ray image (magenta),  and the {\it VLA} 5 GHz image (green).}
\end{figure}

\begin{figure}
\plotfiddle{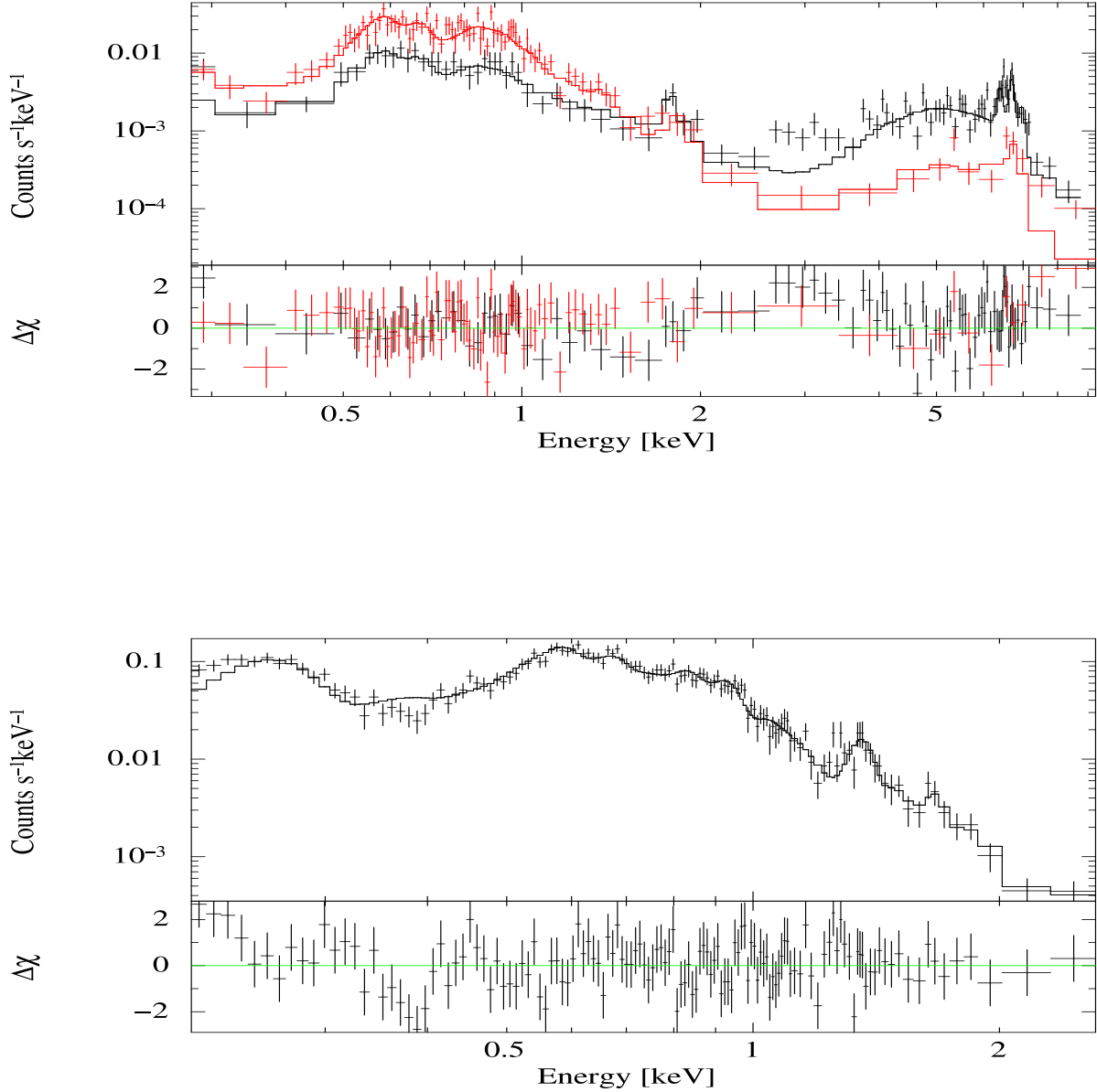}{0.5in}{0}{450}{450}{0}{-300}
\caption{Spectra extracted from the C0, C1, and C3 regions as
identified in Fig.2: (upper panel) spectra 
from central source region, C0, (black), and the inner counter-jet, C1
(red),
with the
corresponding best fits and residuals;
(lower panel) spectrum from the C3 clump and
corresponding best fits and residuals.}
\end{figure}

\end{document}